\documentclass[aps,showpacs,pra,amssymb, amsmath, twocolumn]{revtex4-2}
\usepackage{graphicx}
\usepackage{amssymb}
\usepackage{amsmath}
\usepackage{bm}
\usepackage{xcolor} 
\usepackage{array}
\usepackage{multirow}
\usepackage{tabularx}
\usepackage{booktabs}
\usepackage{textcomp}
\usepackage{mathtools}
\usepackage{bbold}
\usepackage{physics}

\begin{document}
\title{Quantum simulation of Ising spins on Platonic graphs} 
\author{Andrew Byun, Minhyuk Kim, and Jaewook Ahn}
\address{Department of Physics, KAIST, Daejeon 34141, Korea}

\begin{abstract}
\noindent
We present quantum simulation experiments of Ising-like spins on Platonic graphs, which are performed with two-dimensional arrays of Rydberg atoms and quantum-wire couplings. The quantum wires are used to couple otherwise uncoupled long-distance atoms, enabling topology-preserving transformtions of the three-dimensional graphs to the two-dimensional plane. We implement three Platonic graphs, tetrahedron, cube, and octahedron of Platonic solids, and successfully probe their ground many-body spin configurations before and after the quasi-adiabatic control of the system Hamiltonians from the paramagnetic phase to anti-ferromagnetic-like phases. Our small-scale quantum simulations of using less than 22 atoms are limited by experimental imperfections, which can be easily improved by the state-of-the-art Rydberg-atom technologies for more than 1000-atom scales. Our quantum-wire approach is expected to pave a new route towards large-scale quantum simulations.
\end{abstract} 
\maketitle

\section{Introduction} \noindent
Breakthroughs in artificial quantum many-body systems have brought unforeseen opportunities to explore problems in physics, chemistry, material science, and medicine~\cite{quantum simulation review,Monroe2021,Wilkinson2020,Bloch2012,Du2010}. In recent years, Rydberg-atom systems in particular have advanced rapidly, demonstrating quantum-mechanical simulations of the nature of complex quantum materials~\cite{rydberg atom,Saffman2016,Browaeys2016,Browaeys2020,Morgado2021}. There are many advantages to use Rydberg atoms in quantum simulations. Interactions of Rydberg atoms are strong and short-ranged enough to feature the mesoscopic nature of lattice Hamiltonians such as Ising, XY, and XXZ models~\cite{Labuhn2016,Kim2020,Barredo2015,Leseleuc2017,Scholl2021}. Rydberg-atom arrays are easily scalable~\cite{Hyosub2016,Barredo2016,Endres2016} to generate entanglements of a huge number of atoms~\cite{Omran2019}, to investigate many-body dynamics near quantum phase transitions~\cite{crystallization,51 quantum simulation,Lienhard2018,Keesling2019,Bluvstein2021}, and to study topological effects~\cite{Semeghini2021}. Three-dimensional arrangements~\cite{Lee2016,Barredo2018} of Rydberg atoms are used to probe the physical properties of tree lattices~\cite{cayley tree} and are expected to investigate many-body physics of more complex arrangements~\cite{Barredo2018}. In reported experiments~\cite{2d antiferro,Ebadi2021}, the number of Rydberg atoms starts to exceed a few hundred, boosting the expectation towards computational advantages of Rydberg-atom systems in solving combinatorial optimization problems~\cite{Pic2018,Pichler2018,quantum wire exp,Ebadi2022}.

Rydberg-atom technology uses neutral atoms captured and arranged by focused optical beams and coupled with each other through Rydberg-atom dipole blockade interactions~\cite{rydberg atom}. So the inter-atomic distances of coupled Rydberg atoms are lower and upper bounded by the Rayleigh range and Rydberg blockade distance, respectively, which strongly limits geometries and topologies of the qubit couplings of a Rydberg-atom array~\cite{Barredo2018}. One of the latest Rydberg-atom technologies is quantum wiring, the use of auxiliary atoms to couple remote atoms, which enables otherwise impossible, complex qubit coupling networks~\cite{ising quantum wire, quantum wire exp}. In this paper, we use quantum wires to demonstrate topology-preserving transformations of a Rydberg-atom system from a 3D surface to a plane. Three Platonic graphs, the tetrahedron, cube, and octahedron graphs, are experimentally implemented and their many-body ground states are probed via quantum adiabatic controls~\cite{adiabatic quantum computing,crystallization}. 

The rest of the paper is organized as follows. The model Hamiltonian of Rydberg-atom arrays is defined in Sec.~\ref{Model} along with the usages of the quantum wires in our experiments. After we describe the phase diagrams of the Rydberg-atom arrays for the Platonic graphs in Sec.~\ref{Phase} and the experimental procedure in Sec.~\ref{Experiment}, the resulting quantum simulation data of the Platonic graphs are reported in Sec.~\ref{Results}. We conclude by discussing the scaling issues and outlooks in Sec.~\ref{Discussion}.

\section{Quantum wires for Platonic graphs} \label{Model} \noindent
We consider a mathematical graph, $G(V,E)$, to model the interactions of an atom array, in which the vertices, $V$, and edges, $E$, respectively, represent the atoms and pairwise atom-atom interactions. In the Rydberg-atom blockade regime~\cite{rydberg atom} of adjacent atoms, the Hamiltonian $\hat H_G$ of the atom array in $G$ is given by
\begin{equation} \label{H}
\hat H_{G}=\frac{\hbar}{2}\sum_{i\in V} \left(\Omega \hat \sigma_{x,i}-\Delta \hat \sigma_{z,i}\right) +\sum_{(i,j)\in E}  U \hat n_i \hat n_j,
\end{equation}
where $\Omega$ and $\Delta$ are the Rabi frequency and detuning of the optical excitation of the atoms from the ground state, denoted by $\ket{\downarrow}=\ket{0}$, to a Rydberg state, $\ket{\uparrow}=\ket{1}$, and $U=C_6/d^6$ is the van der Waals interaction, of coefficient $C_6$, between the adjacent atoms. The distance of all adjacent atom pairs is kept to be $d$, which is smaller than the blockade distance, $d_R=(C_6/\hbar \Omega)^{1/6}$, i.e., $d<d_R$. $\hat \sigma_{x,z}$ and $\hat n=(\hat \sigma_z+1)/2$ are the operators for Pauli $x,z$ and Rydberg excitation number. This Hamiltonian for quantum Ising spins~\cite{51 quantum simulation, crystallization} has been considered for maximum independent set problems (MIS) of $G$~\cite{Pic2018,Pichler2018,quantum wire exp,Ebadi2022}.

\begin{figure*}[hbt]
\includegraphics[width=2.0\columnwidth]{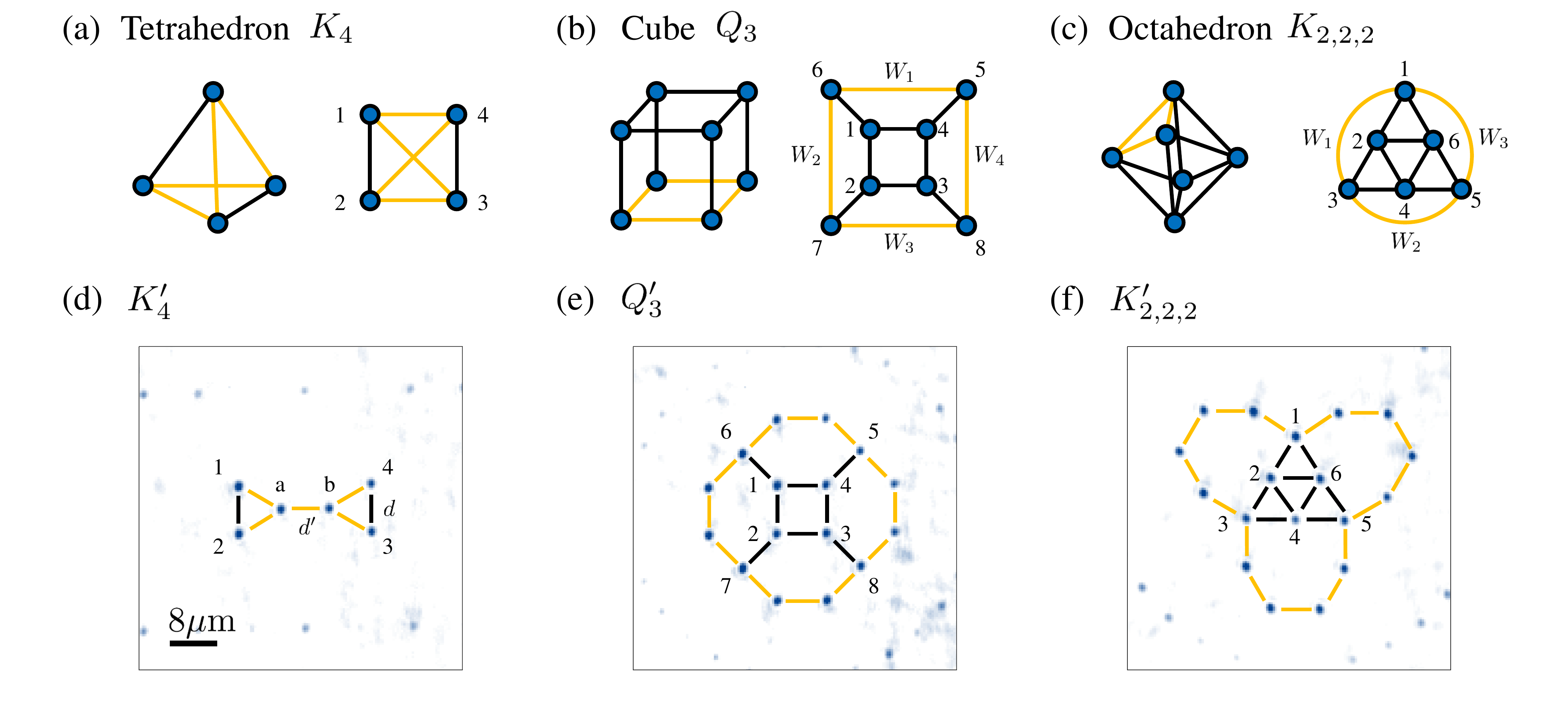}
\caption{Platonic graphs and their two-dimensional arragnements: (a) The tetragonal graph $K_4$,  (b) the cube graph $Q_3$, and (c) the octahedron graph $K_{2,2,2}$, in which the yellow edges are to be replaced by quantum wires. (d) The quantum-wired tetragonal graph, $K_4'$, uses one quantum wire, $W$, of two wire atoms, to replace the four yellow edges of $K_4$, so there are six atoms and five equal-length edges in $K_4'$. (e) The quantum-wired cube graph, $Q_3'$ uses four quantum wires, $W_{1,2,3,4}$, each of two wire atoms, to replace the four yellow edges of $Q_3$, so there are 16 atoms and 20 edges in $Q_3'$. (f) The quantum-wired octahedron graph, $K_{2,2,2}'$ uses three quantum wires, $W_{1,2,3}$, each of four wire atoms, to replace the three yellow edges of $K_{2,2,2}$ so there are 18 atoms and 24 edges in $K_{2,2,2}'$. The lengths are $d'$ for the edges involved with quantum wires and $d$ for all other edges.}
\label{Fig1}
\end{figure*}

Figure~\ref{Fig1} shows, in the first row, three Platonic graphs to be constructed in this work. They are, in the graph theory notations, $K_4$, the tetrahedron graph ($f=4$), $Q_{3}$, the cube graph ($f=6$), and $K_{2,2,2}$, the octahedron graph ($f=8$), as shown in Figs.~\ref{Fig1}(a,b,c), respectively, where $f$ denotes the facet number, i.e., $f=2-||V||+||E||$ in Euler's polyhedron formula. While the given platonic solids are three-dimensional, their graphs are planar, so they are in principle implementable on the plane. However, planar arrangements of atoms for these planar graphs require non-adjacent atoms to be coupled, as if they were adjacent, so we choose to use quantum wires~\cite{quantum wire exp}. 

Quantum wires~\cite{ising quantum wire, quantum wire exp} can be substituted for chosen edges of a graph $G$, in cases when we determine ground-state spin configurations and energies of $H_G$. For the MIS phase which is defined by the parameter conditions, $\Omega=0$ and $0<\Delta<U$, in Eq.~\eqref{H}, the edge of $G$ can be replaced by a quantum wire, which is simply a chain of an even number of wire atoms in an anti-ferromagnetic (AF) spin state. In the MIS phase, in which no adjacent atoms are simultaneously excited to the Rydberg state, the state of the quantum wire, say one with two atoms, must be in either $\ket{00}_W$, $\ket{01}_W$, or $\ket{10}_W$, so the quantum-wire introduces no additional energies~\cite{quantum wire exp}.

\begin{figure*}[t]
\includegraphics[width=1.8\columnwidth]{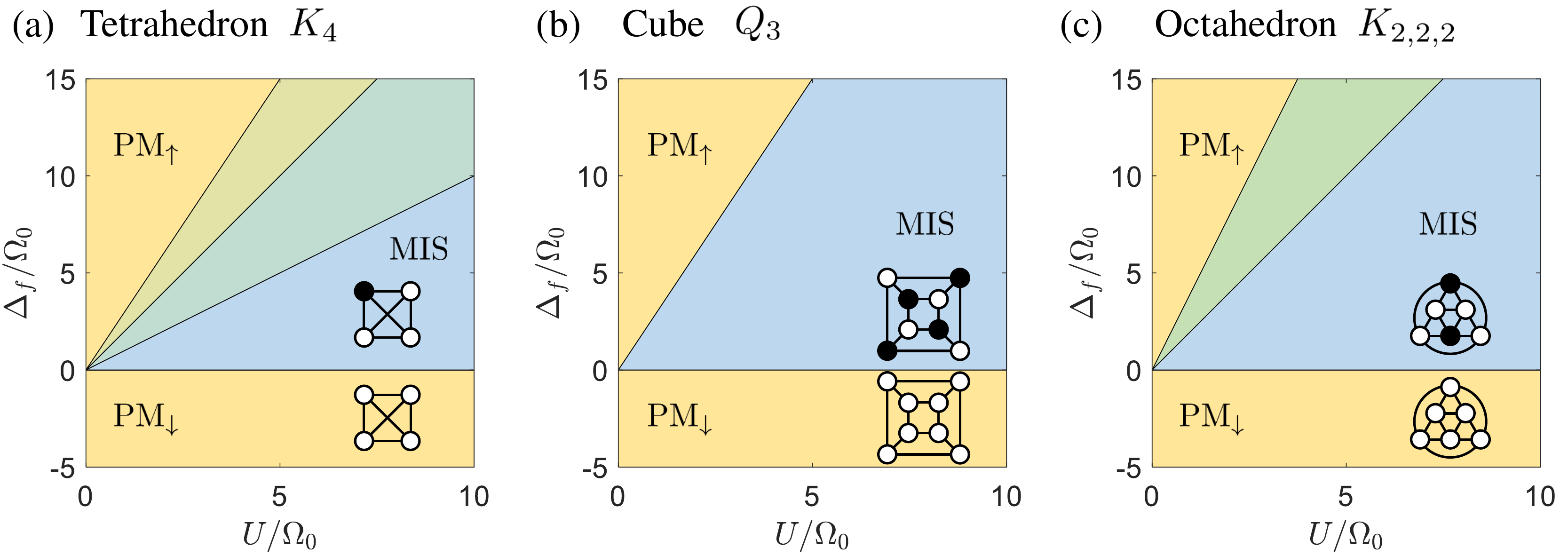}
\caption{Phase diagrams of Platonic graphs: (a) The tetrahedron graph $K_4$, (b) the cube graph $Q_3$, and (c) the octahedron graph $K_{2,2,2}$. There are common paramagnetic phases, PM$_{\downarrow}$ and PM$_{\uparrow}$ in the regions, $\Delta_f<0$ and $\Delta_f>3U$, respectively, and graph-dependent AF-like phases between them. The first AF-like phases above PM$_{\downarrow}$ are the MIS phases of the graphs, in which all adjacent atoms that are denoted by the edges of the graphs are Rydberg blockaded. (a) The MIS phase of $K_4$ allows one Rydberg atom, i.e., the ground many-body state is the superposition of four singly-excited configurations in Eq.~\eqref{MISK4}. The other AF-like phases of $K_4$ are the superpositions of doubly-excited and triply-excited configurations, respectively located in $U<\Delta_f<2U$ and $2U<\Delta_f<3U$. (b) The MIS phase of $Q_3$ is the superposition of two quadruply-excited configurations as in Eq.~\eqref{MISQ3}. (c) The MIS phase of $K_{2,2,2}$ is the superposition of three doubly-excited configurations, as in Eq.~\eqref{MISK222}, and the other AF-like phase is the superposition of three quadruply-excited configurations, located in $2U<\Delta<3U$. Experiments in Sec.~\ref{Experiment} are performed by adiabatically evolving the atom arrays from the PM$_{\downarrow}$ to the MIS phases.
}
\label{Fig2}
\end{figure*}

As an example, let us consider the ground-state spin configurations of the tetrahedron graph, $K_4$, and compare them with the ones of a quantum-wired graph, $K_4'$. As shown in Fig.~\ref{Fig1}(d), our choice of $K_4'$ has two wire atoms, $a$,$b$, which are used to couple the four atoms, 1,2,3,4 of $K_4$. The MIS ground state of the tetrahedron graph, $K_4$, is given by
\begin{equation} \label{MISK4}
\ket{{\rm MIS}(K_4)}=\frac{\ket{1000}+\ket{0100}+\ket{0010}+\ket{0001}}{2}
\end{equation}
which is the four-qubit $W$ state. The MIS ground state of the quantum-wired graph, $K_4'$, is given by
\begin{widetext}
\begin{eqnarray}
\ket{{\rm MIS}(K_4')} &=& \sqrt{\frac{3}{32}} \ket{10}_W \otimes \left(\ket{0001}+\ket{0010}\right) + \sqrt{\frac{3}{32}} \ket{01}_W \otimes \left(\ket{0100}+\ket{1000}\right) , \nonumber \\
&+& \sqrt{\frac{5}{32}} \ket{00}_W  \otimes \left(\ket{0101}+\ket{0110}+\ket{1001}+\ket{1010}\right)
\label{MISK4'}
\end{eqnarray}
\end{widetext}
where the first two terms satisfy the AF condition of the quantum wire but the last term fails. Therefore, if we simply discard the last term, by performing a conditional measurement of $\ket{{\rm MIS}(K_4')}$, given the condition of the quantum-wire's AF configurations, i.e., $\ket{10}_W$ and $\ket{01}_W$ only, we can obtain the probability distribution of $\ket{{\rm MIS}(K_4)}$. With the a priori information of the symmetric phase relation of the graph, we obtain $\left(\bra{01}_W+\bra{10}_W\right) \ket{{\rm MIS}(K_4')} \rightarrow \ket{{\rm MIS}(K_4)}$. 

In general, the MIS ground states of a general graph $G$ and its quantum-wired graph $G'$ are related as
\begin{equation}
\bra{\rm AF}_W \ket{{\rm MIS}(G')} \rightarrow \ket{{\rm MIS}(G)},
\label{G'toG}
\end{equation}
where $\ket{\rm AF}_W$ denotes AF configurations of the quantum wire. In experiments below, we use quantum-wired graphs, $K_4'$ of one quantum wire, $Q_3'$ of four quantum wires, and $K_{2,2,2}'$ of three quantum wires, as shown respectively in Figs.~\ref{Fig2}(d,e,f), to obtain the MIS ground states of the Platonic graphs, $K_4$, $Q_3$, and $K_{2,2,2}$.

\section{Phase diagrams} \label{Phase} \noindent
The phase diagrams, or the ground-state spin configurations, of the Hamiltonian $H_G(U, \Delta_f, \Omega=0)$ are shown in Fig.~\ref{Fig2}  for the chosen Platonic graphs. The Hamiltonian $H_G$ in Eq.~\eqref{H} has two competing energy terms: the laser detuning term which favors Rydberg atoms, or the up spins, and the interaction term which favors no adjacent double excitations. So, there are two paramagnetic phases, PM$_\downarrow$ in the $\Delta<0$ region and PM$_\uparrow$ in $\Delta>3U$, and AF-like phases between them. Among the AF-like phases, we are interested in the MIS phase, which has no adjacent spin pairs at all. The MIS phase regions differ by graphs. Below we consider the MIS phases of the tetrahedron, cube, and octahedron graphs, respectively.

\textit{Tetrahedron}: The tetrahedron graph, $K_4$, has four vertices and each vertex is edged to all others. In the MIS phase of no adjacent Rydberg atoms, only one atom is allowed to be excited to the Rydberg state. So, the MIS ground state is $\ket{\rm{MIS}(K_4)}$ in Eq.~\eqref{MISK4}, which is the super-atom\cite{superatom} superposition state. The MIS phase region, MIS($K_4$), is located in $0<\Delta<U$, as shown in Fig.~\ref{Fig2}(a). Between MIS($K_4$) and PM$_{\uparrow}$, there are two additional AF-like phases in regions $U<\Delta<2U$ and $2U<\Delta<3U$, which have one and two adjacent Rydberg atom pairs, respectively. 

\textit{Cube}: The cube graph, $Q_3$, has eight vertices, each edged to three other vertices. In the MIS phase, MIS($Q_3$), when one atom is in the Rydberg state, its face-diagonal, three vertices are allowed to become Rydberg atoms, so the MIS ground state is 
\begin{equation} \label{MISQ3}
\ket{{\rm MIS}(Q_3)}= \frac{\ket{01010101}+\ket{10101010}}{\sqrt{2}}.
\end{equation}
Any other states, having less or more Rydberg atoms, are not ground states in any region in $0<\Delta<3U$, because Rydberg deexcitation or excitation of $\ket{\rm{MIS}(Q_3)}$ increases the energy by $\Delta>0$ or $3U-\Delta$, respectively. So, the MIS phase is the only AF-like phase of the cube graph, as shown in Fig.~\ref{Fig2}(b).

\textit{Octahedron}: The octahedron graph, $K_{2,2,2}$, has six vertices, each edged to four other vertices. There are two AF-like phases, the MIS phase of two Rydberg atoms, i.e., 
\begin{equation} \label{MISK222}
\ket{{\rm MIS}(K_{2,2,2})}= \frac{\ket{100100}+\ket{010010}+\ket{001001}}{\sqrt{3}}
\end{equation}
and its inversion of four Ryberg atoms. Their equal energy boundary is $\Delta=2U$, so the MIS phase of the octahedron graph is located in $0<\Delta<2U$ and the inverted MIS phase is in $2U<\Delta<3U$, as shown in Fig.~\ref{Fig2}(c).

\section{Experimental procedure} \label{Experiment} \noindent 
The MIS phases of the Platonic graphs, $K_4$, $Q_3$, and $K_{2,2,2}$, are probed with quantum adiabatic control of Hamiltonians, $H_G'$, of the quantum-wired graphs, $G'=K_4'$, $Q_3'$, and $K_{2,2,2}'$. The Hamiltonians are changed from their paramagnetic phase region, PM$_\downarrow$, to the MIS phase region, respectively. Quantum simulation experiments of this kind were previously performed with various Rydberg-atom arrays~\cite{Kim2020,2d antiferro, 51 quantum simulation, cayley tree, quantum wire exp}. 

We used rubidium ($^{87}$Rb) atoms, which were cooled in a magneto-optical trap and optically pumped to the ground state $\ket{0}=\ket{\downarrow}=\ket{5S_{1/2}, F=2, m_F=2}$. The atoms were then captured by far-off resonance dipole traps (optical tweezers) and arranged to positions for $G'$~\cite{rydberg tweezer}. Table~\ref{Tab1} lists the atom positions used in our experiments for $G'=K_4'$, $Q_3'$, and $K'_{2,2,2}$. The nearest-neighbor distances were the same $d=8.0\pm 0.3$~$\mu$m and the blockade distance was $d_B=10.55$~$\mu$m at $\Omega_0/2\pi=0.74$~MHz. The Rydberg state was $\ket{1}=\ket{\uparrow}=\ket{71S_{1/2}, J=1/2, m_J=1/2}$, excited by the two-photon off-resonant transition, $\ket{0}\rightarrow \ket{m} \rightarrow \ket{1}$, via the intermediate state $\ket{m}=\ket{5P_{3/2}, F'=3, m_{F'}=3}$. We used a 780-nm laser field (Toptical DL Pro 780) of $\Omega_{780}=(2\pi)112$~MHz for $\ket{0} \rightarrow \ket{m}$ transition and a 480-nm laser field (Toptica TA-SHG Pro) of $\Omega_{480}=(2\pi)7.4$~MHz for $\ket{m}\rightarrow\ket{1}$, and the intermediate detuning was $\Delta_{m}=(2\pi)560$~MHz. The effective Rabi frequency $\Omega$ was varied up to the max $\Omega_0=\Omega_{780}\Omega_{480}/2\Delta_{m}=(2\pi)0.74$~MHz.  

\begin{table}
\caption{Atom positions of the experimental graphs}
\begin{ruledtabular}
\begin{tabular}{@{}lllll@{}}
Graphs & \multicolumn{4}{l}{Atom positions $(x,y)$ [$\mu$m]}   \\
\hline\hline
\multirow{3}{*}{Tetrahedron $K'_4$} & 1: &(-10.9, 4.0) & 2: &(-10.9, -4.0) \\
& 3:& (10.9, -4.0) & 4:& (10.9, 4.0)  \\
& $W$: & ($\pm$ 4.0, 0.0) && \\ 
\hline
\multirow{6}{*}{Cube $Q'_3$}  & 1:& (-4.0, 4.0) & 2:& (-4.0, -4.0) \\ 
& 3:& (4.0, -4.0) & 4:& (4.0, 4.0) \\
& 5:& (9.7, 9.7)  & 6:& (-9.7, 9.7)\\
& 7:& (-9.7, -9.7) & 8:& (9.7, -9.7) \\
& $W_1$:& ($\pm$ 4.0, 15.3) & $W_2$: &(-15.3, $\pm$ 4.0) \\
& $W_3$: &(-4.0, $\pm$ 15.3) & $W_4$:  &(15.3, $\pm$ 4.0) \\
\hline
\multirow{3}{*}{Octahedron $K_{2,2,2}'$}  & 1:& (0.0, 9.8) & 2:& (-4.0, 2.9) \\
& 3:& (-8.0, -4.0) & 4:& (0.0, -4.0)\\
& 5:& (8.0, -4.0) &6:& (4.0, 2.9) \\
& $W_1$:& \multicolumn{3}{l}{\begin{tabular}{ll} (-6.9, 13.8) & (-14.9, 13.8) \\ (-18.9, 6.9) & (-14.9, 0.0) \end{tabular}} \\
& $W_2$:& \multicolumn{3}{l}{\begin{tabular}{ll} (-8.0, -12.0) & (-4.0, -18.9) \\ (4.0, -18.9) & (8.0, -12.0) \end{tabular}} \\
& $W_3$:& \multicolumn{3}{l}{\begin{tabular}{ll} (14.9, 0.0) & (18.9, 6.9) \\ (14.9, 13.8) & (6.9, 13.8) \end{tabular}} \\
\end{tabular}
\end{ruledtabular}
\label{Tab1}
\end{table}

The atoms were initially in the PM$_\downarrow$ phase, i.e., $\ket{\Psi(t=0)}=\ket{0}^{N'}$, where $N'$ the number of atoms in $G'$, and then adiabatically driven to their MIS phase, $\ket{\Psi(t=t_f)}=\ket{{\rm MIS}(G')}$. The control parameters of $H_{G'}$ in Eq.~\eqref{H} were changed in three stages. In the first stage, the Rabi frequency $\Omega(t)$ was linearly tuned on from $\Omega(t=0)=0$ to $\Omega(t_1)=\Omega_0$ and the detuning $\Delta(t)$ was maintained at $\Delta(t=0-t_1)=-3$~MHz. In the second stage, we maintain the Rabi frequency $\Omega(t=t_1-t_2)=\Omega_0$ and linearly change the detuning to $\Delta(t_2)=\Delta_f$. In the final stage, the Rabi frequency was linearly changed to zero, i.e., $\Omega(t_f)=0$, and the detuning was maintained at $\Delta(t=t_2-t_f)=\Delta_f$. We used $t_f=4.0$~$\mu$s, $t_1=t_f/10$, $t_2=t_f-t_1$, and $\Delta_f=2.0$~MHz for tetrahedron experiment and 3.0~MHz for cube and octahedron experiments.  $\Omega(t)$ and $\Delta(t)$ were changed with a radio-frequency programmable synthesizer (Moglabs XRF) and acousto-optic modulators~(AOM)~\cite{cayley tree}. The final Hamiltonian parameters are $(U/\Omega_0, \Delta_f/\Omega_0)=(5.27, 2.70)$ for the tetragonal $K_4$ and $(5.27, 4.05)$ for the cube $Q_3$ and octahedron $K_{2,2,2}$ experiments, which are located in their respective MIS phases in Fig.~\ref{Fig3}.

After the atoms were driven to the MIS phase, their final states were measured by collecting the fluorescence images of the cyclic transition, $\ket{5S_{1/2}, F=2} \leftrightarrow \ket{5P_{3/2},F'=3}$ in an electron multiplying charge coupled device (Andor iXon Ultra 897). We repeated the above measurement by 927 times for $K_4$, 3292 for $Q_3$, and 1574 for $K_{2,2,2}$ to obtain their probabilities of all spin configurations.

\section{Results}\label{Results} 
\begin{figure*}[t]
\includegraphics[width=2.0\columnwidth]{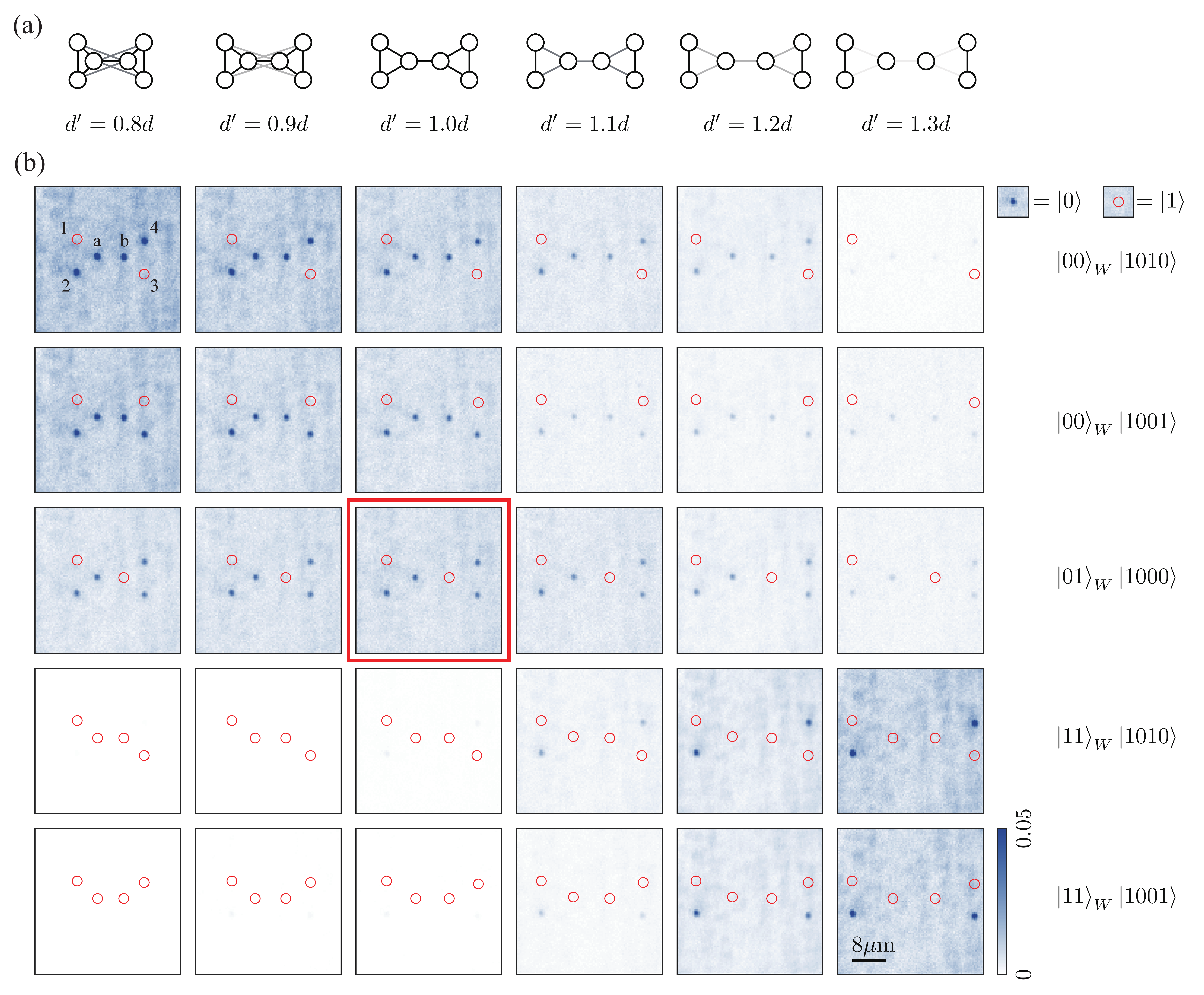}
\caption{$K_4'$-like graphs of two types of edges: (a) Six different graphs are constructed with $d'/d=0.8$, 0.9, 1.0, 1.1, 1.2, and 1.3 (from left to right), among which the $d'=d$ graph corresponds to $K_4'$. (b) Fluorescence images of the atom arrays driven to $H(\Omega_f=0,\Delta_f=2.0\rm{MHz}, U=3.9\rm{MHz})$ and sorted based on their spin configurations. Six characteristic spin configurations are shown, (from top to bottom) $\ket{00}_{W}\ket{1010}$, $\ket{00}_{W}\ket{1001}$, $\ket{01}_{W}\ket{1000}$, $\ket{11}_{W}\ket{1010}$, and $\ket{11}_{W}\ket{1001}$. The scale bar represents the relative probabilities of the configurations per graph. The numbers of collected events are 1057, 971, 927, 906, 2866, and 918, respectively, for the graphs.}
\label{Fig3}
\end{figure*}

\noindent
Before we perform the quantum simulation of the platonic graphs, we first test the working principle of the quantum wire by using an example of $K_4'$ and varying the edge distance of the quantum-wired graph. We use two kinds of edges, $d_{12}=d_{34}=d$ and $d_{1a}=d_{2a}=d_{ab}=d_{b3}=d_{b4}=d'$, which are black- and yellow-colored edges, respectively, in Fig.~\ref{Fig1}(d), where $a$,$b$ denote the two wire atoms and $d'$ is the length of the edges involved with the wire atoms. We use six different atom-arrays of $d'$ varied from $0.8d$ to $1.3d$ as in Fig.~\ref{Fig3}(a) (from the left to the right). It is noted that, as $d'/d$ increases from zero to $\infty$, the six-atom system changes from a super-atom to a pair of isolated dimers, and the $d'=d$ case corresponds to the $K_4'$ graph in the third column in Fig.~\ref{Fig3}.

In Fig.~\ref{Fig3}(b), we show configuration-dependent fluorescence images of the atoms adiabatically driven to the MIS-phase condition,
i.e., $0<\Delta_f/U=0.51<1$. There are six characteristic spin-configurations, which are, from the top to the bottom, $\ket{00}_{W}\ket{1010}$, $\ket{00}_{W}\ket{1001}$, $\ket{01}_{W}\ket{1000}$, $\ket{11}_{W}\ket{1010}$, and $\ket{11}_{W}\ket{1001}$. Other spin-configurations of the same rotational or reflection symmetries are observed similarly. The contrasts of the images in Fig.~\ref{Fig2}(b) are normalized per column to compare their relative occurrences among spin-configurations and graphs. In the first column for $d'/d=0.8$, the most-strongly interacting six atoms, the most frequently observed configuration is $\ket{00}_{W}\ket{1010}$, which has both the wire atoms in the ground state. In the last column for $d'/d=1.3$, the pair of isolated dimers, $\ket{11}_{W}\ket{1001}$ is the most frequently observed configuration, in which both the wire atoms are Rydberg atoms. Both of these cases, $\ket{00}_{W}\ket{1010}$ and 
$\ket{11}_{W}\ket{1001}$ fail the AF quantum-wire condition. On the other hand, in the third column, which corresponds  $K_4'$ ($d'=d$), the highlighted $\ket{01}_{W}\ket{1000}$ configuration is significantly observed, which satisfies the AF quantum-wire condition. The other two significantly observed configurations are $\ket{00}_{W}\ket{1010}$ and $\ket{00}_{W}\ket{1001}$, agreeing well with the expected ground many-body state of $K_4'$ in Eq.~\eqref{MISK4'}. 

\begin{figure*}[t]
\includegraphics[width=2.0\columnwidth]{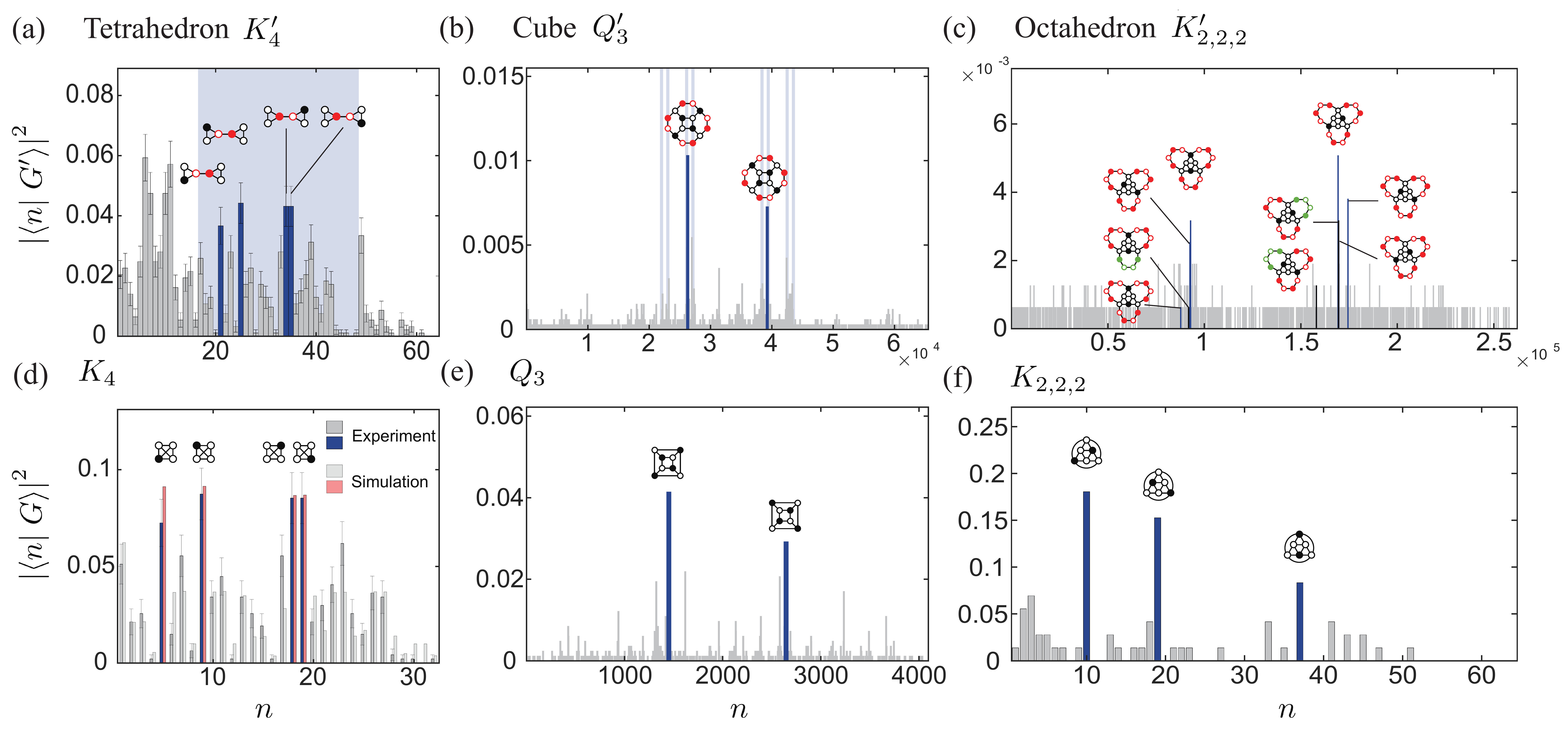}
\caption{Probability distributions, $|\langle n \ket{G'}|^2$, of quantum-wired Platonic graphs driven to their MIS phases: (a) $G'=K_4'$, (b) $Q_3'$, and (c) $K_{2,2,2}'$. The numbers of collected experimental events are 927, 3292, and 1574, respectively. (d-f) After the AF conditions of the quantum wires are imposed, there remain 469, 820, and 72 events, respectively, which are used to reconstuct the probability distributions, $|\langle n \ket{G}|^2$, of the Platonic graphs, (d) $G=K_4$, (e) $Q_3$, and (f) $K_{2,2,2}$. In (d), a numerical simulation of the evolved dynamics of (a) is shown in comparison.}
\label{Fig4}
\end{figure*}

Main results of the quantum simulation experiments of the platonic graphs are summarized in Fig.~\ref{Fig4}. The atom arrays of the quantum-wired graphs, $G'=K_4'$, $Q'_3$, and $K'_{2,2,2}$, are adiabatically driven from the PM$_\downarrow$ phase to their MIS phases and their probabilities are plotted per spin configurations, before and after the AF quantum-wire conditions are imposed,  in Figs.~\ref{Fig4}(a-c) and Figs.~\ref{Fig4}(d-e), respectively. The experimental probability distributions of the platonic graphs, $K_4$, $Q_3$, and $K_{2,2,2}$ are in good agreements with what are expected from their MIS ground states, $\ket{{\rm MIS}(K_4)}$, $\ket{{\rm MIS}(Q_3)}$, and $\ket{{\rm MIS}(K_{2,2,2})}$, respectively in Eqs.~\eqref{MISK4}, \eqref{MISQ3}, and \eqref{MISK222}, as explained below.

\textit{Tetrahedron}: In Fig.~\ref{Fig4}(a), we plot the experimental $K_4'$ probabilities, $|\langle n \ket{{\rm Exp}(K_4')}|^2$, where $n$ is a decimal number enumerating the spin configuration of the $K_4'$ atom-array of two wire atoms and four $K_4$ atoms in such a way that $\ket{n=1}=\ket{00}_W\otimes\ket{0000}$, $\ket{2}=\ket{00}_W\otimes\ket{0001}$, $\cdots$, $\ket{64}=\ket{11}_W\otimes\ket{1111}$. The shaded region in Fig.~\ref{Fig3}(a) corresponds to the spin configurations satisfying the AF condition of the quantum wire. The max-population peaks in the AF region are identified to be $\ket{21}=\ket{01}_W\ket{0100}$, $\ket{25}=\ket{01}_W\ket{1000}$, $\ket{34}=\ket{10}_W\ket{0001}$, and $\ket{35}=\ket{10}_W\ket{0010}$, agreeing well with the spin configurations of $\ket{{\rm MIS} (K_4)}$ in Eq.~\eqref{MISK4}. It is noted that the significant populations in the first quadrant in Fig.~\ref{Fig3}(b), which corresponds to $\ket{00}_W$, are the four spin-configurations in the third term of Eq.~\eqref{MISK4'}, so they violate the AF condition. In Fig.~\ref{Fig4}(d), after the AF quantum-wire condition is imposed, we plot the probability distribution of $\ket{{\rm Exp}(K_4)}$, which is the normalization of $|\bra{{\rm AF}}_W \ket{{\rm Exp}(K_4')}|^2$, and compared it with a numerical simulation. The quantum evolution of the $K_4'$ atoms is numerically traced along the adiabatic control path introduced in Sec.~\ref{Experiment} with a Lindbladian master equation~\cite{Lindblad} which takes into account experimental noises from laser phase noises and bit-flip detection errors of $P_{0\rightarrow1'}=0.12$, $P_{1\rightarrow0'}=0.09$~\cite{noise, noise few atom}. The experimentally observed $K_4$ probability, which is the sum of the populations of the four peaks, $\ket{5}=\ket{01}_W\ket{0100}$, $\ket{9}=\ket{01}_W\ket{1000}$, $\ket{18}=\ket{10}_W\ket{0001}$, and $\ket{19}=\ket{10}_W\ket{0010}$, is $|\langle {\rm MIS}(K_4)\ket{{\rm Exp(K_4)}}|^2=0.33\pm0.03$.

\textit{Cube}: The experimental observed $Q_3'$ probability distribution of $|\langle n \ket{{\rm Exp}(Q_3')}|^2$, for $n=1,\cdots,2^{16}$, are plotted in Fig.~\ref{Fig4}(b). In the MIS phase of the $Q_3'$ graph, we expect
\begin{eqnarray}
&&\ket{{\rm MIS}(Q_3')} = (\ket{01}_{W_1}\ket{10}_{W_2}\ket{01}_{W_3}\ket{10}_{W_4} \ket{10101010} \nonumber\\
&& \quad +\ket{10}_{W_1}\ket{01}_{W_2}\ket{10}_{W_3}\ket{01}_{W_4}\ket{01010101})/\sqrt{2}, 
\end{eqnarray}
because the state of each wire-atom pair which couples a pair of $Q_3$ atoms in $\ket{01}$ ($\ket{10}$) is determined to be $\ket{10}_W$ ($\ket{01}_W$). We observed two max-populated spin-configurations, $\ket{26283}$ and $\ket{39254}$, in Fig.~\ref{Fig4}(b), agreeing well with the two spin configurations in $\ket{{\rm MIS}(Q_3')}$. In Fig.~\ref{Fig4}(e), after imposing the AF conditions of the quantum wires, we plot the experimental $Q_3$ probability distribution of $\ket{{\rm Exp}(Q_3)}$, of which the result agrees well with $\ket{{\rm MIS}(Q_3)}$ in Eq.~\eqref{MISQ3}. The experimentally observed probability of $Q_3$ is $|\langle {\rm MIS}(Q_3)\ket{{\rm Exp}(Q_3)}|^2=0.07\pm 0.01$, which is the sum of populations of $\ket{1451}$ and $\ket{2646}$ in Fig.~\ref{Fig4}(e).

\textit{Octahedron}: In Fig.~\ref{Fig4}(c), we plot $|\langle n \ket{{\rm Exp}(K_{2,2,2}')}|^2$, the experimental probability distribution of $K_{2,2,2}'$ which has twelve quantum-wire atoms and six $K_{2,2,2}$ atoms. The MIS ground state of the $K_{2,2,2}'$ is a little complicated. Among the three quantum wires, each of which has four wire atoms, as in Fig.~\ref{Fig1}(c), two quantum wires, say $W_1$ and $W_2$ for the case when atoms $3$,$6$ are Rydberg atoms, must be in the AF states, $\ket{W_1}=\ket{1010}$ and $\ket{W_2}=\ket{0101}$, but then the other quantum wire, $W_3$, can be in $\ket{1001}$ as well as in $\ket{0101}$ and $\ket{1010}$, because they contribute the same energies. So, the MIS ground state of $K_{2,2,2}'$ is given by
\begin{widetext}
\begin{eqnarray} \label{MISQ3'}
\ket{{\rm MIS}(K_{2,2,2}')} &=& \ket{1010}_{W_1} \ket{0101}_{W_2}\left(a(\ket{0101}+\ket{1010})+b\ket{1001}\right)_{W_3}\ket{001001} \nonumber \\
&+& \left(a(\ket{0101}+\ket{1010})+b\ket{1001}\right)_{W_1} \ket{1010}_{W_2}\ket{0101}_{W_3} \ket{010010} \nonumber \\
&+&  \ket{0101}_{W_1}\left(a(\ket{0101}+\ket{1010})+b\ket{1001}\right)_{W_2}\ket{1010}_{W_3}\ket{100100},
\end{eqnarray}
\end{widetext}
where $a=\sqrt{20/243}$ and $b=\sqrt{41/243}$ are normalization factors. Due to the limited statistics of the $K_{2,2,2}'$ experiment, it is unclear if there are 9 peaks in Fig.~\ref{Fig4}(c), as expected from Eq.~\ref{MISQ3'}; however, after summing the events with the AF quantum-wire conditions, we observe three clear peaks in Fig.~\ref{Fig4}(f), which correspond to the spin-configurations of the MIS phase of $K_{2,2,2}$, agreeing well with $\ket{{\rm MIS}(K_{2,2,2})}$. Experimentally observed $K_{2,2,2}$ probability is $|\langle {\rm MIS}(K_{2,2,2})\ket{{\rm Exp}(K_{2,2,2})}|^2=0.39\pm0.07$, which is the sum of the populations of $\ket{10}=\ket{001001}$ and $\ket{19}=\ket{010010}$, and $\ket{37}=\ket{100100}$.

\section{Discussions and Outlook} \label{Discussion} 
\begin{figure}[t]
\includegraphics[width=1.0\columnwidth]{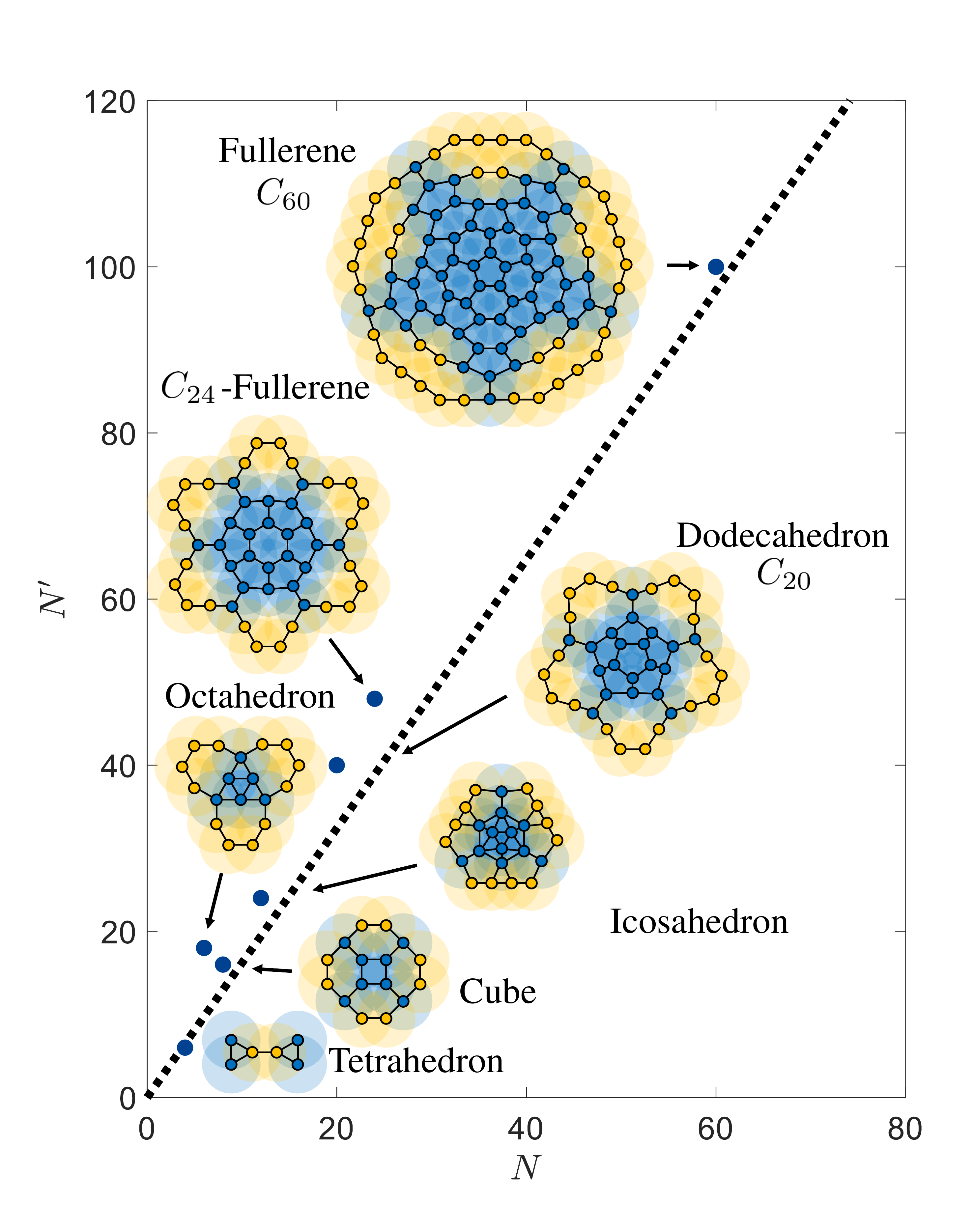}
\caption{Atom-number scaling of $N'$, of a quantum-wire graph $G'$, versus $N$, of a target graph $G$}
\label{Fig5}
\end{figure}

\noindent
Now we turn our attention to the scaling issue involved with the current method of graph transformations from 3D surfaces to 2D planes.
Our experiment presented the transformation of the tetrahedron graph $K_4$ of $N=4$ atoms to the quantum-wired graph $K_4'$ of total $N'=6$ atoms, the cube graph $Q_3$ of $N=8$ to $Q_3'$ of $N'=16$, and the octahedron graph $K_{2,2,2}$ of $N=6$ to $K_{2,2,2}'$ of $N'=22$. In Fig.~\ref{Fig5}, we plot $N$ versus $N'$ for these three Platonic graphs and we also plot other examples of the remaining Platonic graphs, the icosahedron and dodecahedron graphs, and two Fullerene graphs, $C_{24}$ and $C_{60}$. We observe a linear relation between $N$ and $N'$, of which the scaling can be understood as follows: When we move the vertices from the sphere of radius $R$ to the square plane of length $L$, we expect $L \sim R$, because the distances of vertices along a chosen circumference of the sphere can be chosen to be unchanged. The number $N$ of a graph $G$ scales with $N \sim R^2/r_B^2$ and $N'$ of $G'$ scales with $N'\sim L^2/r_B^2$. So, the quantum-wired graph $G'$ can be constructed with $N'\sim N$ to transform $G$ from 3D to 2D.

While the number of atoms required for the 3D-to-2D transformation is scaled favorably, the actual $N'$ in our quantum-wired graph experiments is limited by, for example, the rearrangement probabiliy and bit-flip errors. When $N'$ atoms are to be rearranged to chosen positions, the rearrangement probability of $N'$ atoms is given by $P_r=p^{N'}$, where $p=\exp(-t/t_0)=0.97$ is the single-atom survival probability for an average trap lifetime of $t_0=16$~s and rearrangement time of $t=0.6$~s, which gives $N'<80$ for $P_r>0.01$. The bit-flip errors, due to imperfections of the state preparation and measurements as well as quasi-adiabatic controls, result in statistical ambiguities of the ground-state probability, $P_g'\sim (1-P_{0\rightarrow 1})^{N'/2}(1-P_{1\rightarrow 0})^{N'/2} P_g$, and the probabilites of other states, $P_{\rm others} \sim P_{0\rightarrow 1}(1-P_{0\rightarrow 1})^{N'/2-1}(1-P_{1\rightarrow 0})^{N'/2}P_g$. A criterion of $|P_g'-P_{\rm others}| \sim \sqrt{P_g'(1-P_g')/M}$ demands about $M=100$ repeated experiments, which takes $\sim$ 2 minites for $N'=25$, about $M=2\times 10^5$ experiments or $\sim 3$ days for $N'=80$. Both the considered experimental limitations can be improved by the state of the art Rydberg-atom technologies. One approach to increase the rearrangement probability is using a cryogenic environment of Rydberg atoms~\cite{Schymik2021}, in which the the trap lifetime is measured to be $t_0=6\times 10^3$~s. It is expected that the single-atom survival probability $p=0.9999$ in the cryogenic setup can achieve $N'>1000$ for $P_r = 0.9$. The bit-flip error are significantly reduced with alkaline-earth atomic systems~\cite{Endres2020}, where $P_{0\rightarrow 1}$ and $P_{1\rightarrow }$ are less than 0.01.

In summary, we have performed quantum simulation of Ising-like spins on three Platonic graphs, the tetrahedron, cube, and octahedron graphs, to experimentally obtain their many-body ground states in antiferromagnetic phases. While these graphs are 3D structures due to their coupling structures, we used Rydberg atom wires to transform them to 2D structures, in which the quantum wires adjusted the coupling strengths of stretched edges to be the same as those of the unstretched edges. With the 2D quantum-wired graphs, we control the atom-array Hamiltonians adiabatically from their paramagnetic phase to the AF phases and obtained their many-body AF ground-state spin configurations. The results are in good agreements with the symmetry analysis and numerical simulations. In our estimations, our approach of using quantum wires to transform 3D planar-graph structures could achieve, in conjunction with alkaline-earth atom technologies in cryogenic environments, more than 1000-atom quantum simulations.


\begin{thebibliography}{1}
\bibitem{quantum simulation review} I. M. Georgescu, S. Ashhab, and F. Nori, ``Quantum simulation,'' Rev. Mod. Phys. {\bf 86}, 153 (2014).


\bibitem{Monroe2021} C. Monroe, W. C. Campbell, L.-M. Duan, Z. -X. Gong, A. V. Gorshkov, P. W. Hess, R. Islam, K. Kim, N. M. Linke, G. Pagano, P. Richerme, C. Senko, and N. Y. Yao, ``Programmable quantum simulations of spin systems with trapped ions," Rev. Mod. Phys. {\bf 93}, 025001 (2021).

\bibitem{Wilkinson2020} S. A. Wilkinson and M. J. Hartmann, ``Superconducting quantum many-body circuits for quantum simulation and computing," Appl. Phys. Lett. {\bf 116}, 230501 (2020)

\bibitem{Bloch2012} I. Bloch, J. Dalibard, and S. Nascimbene, ``Quantum simulations with ultracold quantum gases," Nat. Phys. {\bf 8}, 267 (2012).

\bibitem{Du2010} J. Du, N. Xu, X. Peng, P. Wang, S. Wu, and D. Lu, ``NMR Implementation of a Molecular Hydrogen Quantum Simulation with Adiabatic State Preparation," Phys. Rev. Lett. {\bf 104}, 030502 (2010).

\bibitem{rydberg atom} M. Saffman, T. G. Walker, and K. M{\o}lmer, ``Quantum information with Rydberg atoms,'' Rev. Mod. Phys. {\bf 82}, 2313 (2010).

\bibitem{Saffman2016} M. Saffman, ``Quantum computing with atomic qubits and Rydberg interactions: progress and challenges," J. Phys. B: At. Mol. Opt. Phys. {\bf 49} 202001 (2016).

\bibitem{Browaeys2016} A. Browaeys, D. Barredo, and T. Lahaye, ``Experimental investigations of dipole–dipole interactions between a few Rydberg atoms," J. Phys. B: At. Mol. Opt. Phys. {\bf 49} 152001 (2016).

\bibitem{Browaeys2020} A. Browaeys and T. Lahaye, ``Many-body physics with individually controlled Rydberg atoms," Nat. Phys. {\bf 16}, 132-142 (2020).

\bibitem{Morgado2021} M. Morgado and S. Whitlock, ``Quantum simulation and computing with Rydberg-interacting qubits,'' AVS Quantum Sci. {\bf 3}, 023501 (2021).



\bibitem{Labuhn2016} H. Labuhn, D. Barredo, S. Ravets, S. de L\'es\'eleuc, T. Macr\`i, T. Lahaye, and A. Browaeys, ``Tunable two-dimensional arrays of single Rydberg atoms for realizing quantum Ising models," Nature {\bf 534}, 667–670 (2016).


\bibitem{Kim2020} M. Kim, Y. Song, J. Kim, and J. Ahn, ``Quantum-Ising Hamiltonian programming in trio, quartet, and sextet qubit systems,'' PRX Quantum {\bf 1}, 020323 (2020).

\bibitem{Barredo2015} D. Barredo, H. Labuhn, S. Ravets, T. Lahaye, A. Browaeys, and C. S. Adams, ``Coherent Excitation Transfer in a Spin Chain of Three Rydberg Atoms," Phys. Rev. Lett. {\bf 114}, 113002 (2015).

\bibitem{Leseleuc2017} S. de L\'es\'eleuc, D. Barredo, V. Lienhard, A. Browaeys, and T. Lahaye, ``Optical Control of the Resonant Dipole-Dipole Interaction between Rydberg Atoms," Phys. Rev. Lett. {\bf 119}, 053202 (2017).


\bibitem{Scholl2021} P. Scholl, H. J. Williams, G. Bornet, F. Wallner, D. Barredo, T. Lahaye, A. Browaeys, L. Henriet, A. Signoles, C. Hainaut, T. Franz, S. Geier, A. Tebben, A. Salzinger, G. Z\"urn, and M. Weidem\"uller, ``Microwave-engineering of programmable XXZ Hamiltonians in arrays of Rydberg atoms," arXiv:2107.14459 (2021).


\bibitem{Hyosub2016} H. Kim, W. Lee, H.-G. Lee, H. Jo, Y. Song, and J. Ahn, ``In situ single-atom array synthesis by dynamic holographic optical tweezers," Nat. Commun. {\bf 7}, 13317 (2016).


\bibitem{Barredo2016} D. Barredo, S. de L\'es\'eleuc, V. Lienhard, T. Lahaye, and A. Browaeys, ``An atom-by-atom assembler of defect-free arbitrary 2D atomic arrays," Science {\bf 354}, 1021 (2016).


\bibitem{Endres2016} M. Endres, H. Bernien, A. Keesling, H. Levine, E. R. Anschuetz, A. Krajenbrink, and M. D. Lukin, ``Atom-by-atom assembly of defect-free one-dimensional cold atom arrays," Science {\bf 354}, 1024 (2016).


\bibitem{Omran2019} A. Omran, H. Levine, A. Keesling, G. Semeghini, T. T. Wang, S. Ebadi, H. Bernien, A. S. Zibrov, H. Pichler, S. Choi, J. Cui, M. Rossignolo, P. Rembold,S. Montangero, T. Calarco, M. Endres, M. Greiner, V. Vuleti\'c, and M. D. Lukin, ``Generation and manipulation of Schr\"odinger cat states in Rydberg atom arrays," Science {\bf 365}, 570 (2019).




\bibitem{crystallization} P. Schau\ss, J. Zeiher, T. Fukuhara, S. Hild, M. Cheneau, T. Macr\`i, T. Pohl, I. Bloch, C. Gross, ``Crystallization in Ising
quantum magnets," Science, {\bf 347}, 6229, 1455-1458 (2015).

\bibitem{51 quantum simulation} H. Bernien, S. Schwartz, A. Keesling, H. Levine, A. Omran, H. Pichler, S. Choi, A. S. Zibrov, M. Endres, M. Greiner, V. Vuleti\'c and M. D. Lukin ,``Probing many-body dynamics on a 51-atom quantum simulator," Nature {\bf 551}, 579-584 (2017).


\bibitem{Lienhard2018}  V. Lienhard, S. de L\'es\'eleuc, D. Barredo, T. Lahaye, A. Browaeys, M. Schuler, L.-P. Henry, and A. M. L\"auchli, ``Observing the Space- and Time-Dependent Growth of Correlations in Dynamically Tuned Synthetic Ising Models with Antiferromagnetic Interactions," Phys. Rev. X {\bf 8}, 021070 (2018).

\bibitem{Keesling2019} A. Keesling, A. Omran, H. Levine, H. Bernien, H. Pichler, S. Choi, R. Samajdar, S. Schwartz, P. Silvi, S. Sachdev, P. Zoller, M. Endres, M. Greiner, V. Vuleti\'c, and M. D. Lukin, ``Quantum Kibble–Zurek mechanism and critical dynamics on a programmable Rydberg simulator," Nature {\bf 568}, 207-211 (2019).




\bibitem{Bluvstein2021} D. Bluvstein, A. Omran, H. Levine, A. Keesling, G. Semeghini, S. Ebadi, T. T. Wang, A. A. Michailidis, N. Maskara, W. W. Ho, S. Choi, M. Serbyn, M. Greiner, V. Vuletic, M. D. Lukin, ``Controlling quantum many-body dynamics in driven Rydberg atom arrays," Science {\bf 371}, 6536 (2021).

\bibitem{Semeghini2021} G. Semeghini, H. Levine, A. Keesling, S. Ebadi, T. T. Wang, D. Bluvstein, R. Verresen, H. Pichler, M. Kalinowski, R. Samajdar, A. Omran, S. Sachdev, A. Vishwanath, M. Greiner, V. Vuleti\'c, and M. D. Lukin, ``Probing topological spin liquids on
a programmable quantum simulator," Science {\bf 374}, 1242 (2021).

\bibitem{Lee2016} W. Lee, H. Kim, and J. Ahn, ``Three-dimensional rearrangement of single atoms using actively controlled optical microtraps," Opt. Express {\bf 24}(9), 9816 (2016).

\bibitem{Barredo2018} D. Barredo, V. Lienhard, S. L\'es\'eleuc, T. Lahaye, and A. Browaeys,``Synthetic three-dimensional atomic structures assembled atom by atom," Nature {\bf 561}, 79–82 (2018).




\bibitem{cayley tree} Y. Song, M. Kim, H. Hwang, W. Lee, and J. Ahn, ``Quantum simulation of Cayley-tree Ising Hamiltonians with three-dimensional Rydberg atoms,'' Phys. Rev. Research. {\bf 3}, 013286 (2021).



\bibitem{2d antiferro} P. Scholl, M. Schuler, H. J. Williams, A. A. Eberharter, D. Barredo, K-N. Schymik, V. Lienhard, L-P. Henry, T. C. Lang, T. Lahaye, A. M. L\"auchli, and A. Browaeys,``Quantum simulation of 2D antiferromagnets with hundreds of Rydberg atoms," Nature {\bf 595}, 233–238 (2021).

\bibitem{Ebadi2021} S. Ebadi, T. T. Wang, H. Levine, A. Keesling, G. Semeghini, A. Omran, D. Bluvstein, R.Samajdar, H. Pichler, W. W. Ho, S. Choi, S. Sachdev, M. Greiner, V. Vuleti\'c, and M. D. Lukin,``Quantum Phases of Matter on a 256-Atom Programmable Quantum Simulator," Nature {\bf 595}, 227–232 (2021).


\bibitem{Pic2018} H. Pichler, S.-T. Wang, L. Zhou, S. Choi, and M. D. Lukin, ``Quantum optimization for maximum independent set using Rydberg atom arrays,'' arXiv:1808.10816 (2018).

\bibitem{Pichler2018} H. Pichler, S.-T. Wang, L. Zhou, S. Choi, and M. D. Lukin, ``Computational complexity of the Rydberg blockade in two dimensions," arXiv:1809.04954 (2018).

\bibitem{quantum wire exp} M. Kim, K. Kim, J. Hwang, E-G. Moon, and J. Ahn, ``Rydberg Quantum Wires for Maximum Independent Set Problems with Nonplanar and High-Degree Graphs," arXiv:2109.03517

\bibitem{Ebadi2022} S. Ebadi, A. Keesling, M. Cain, T. T. Wang, H. Levine, D. Bluvstein, G. Semeghini, A. Omran, J. Liu, R. Samajdar, X.-Z. Luo, B. Nash, X. Gao, B. Barak, E. Farhi, S. Sachdev, N. Gemelke, L. Zhou, S. Choi, H. Pichler, S. Wang, M. Greiner, V. Vuleti\'c, M. D. Lukin, ``Quantum Optimization of Maximum Independent Set using Rydberg Atom Arrays," arXiv:2202.09372







\bibitem{ising quantum wire} X. Qiu, P. Zoller, and X. Li, ``Programmable Quantum Annealing Architectures with Ising Quantum Wires," PRX Quantum {\bf1}, 020311 (2020).



\bibitem{adiabatic quantum computing} T. Albash and D. A. Lidar, ``Adiabatic quantum computation ," Rev. Mod. Phys. {\bf 90}, 015002 (2018).



\bibitem{rydberg tweezer} A. Browaeys and T. Lahaye, ``Many-body physics with individually controlled Rydberg atoms," Nat. Phys. {\bf16}, 132–142 (2020).



\bibitem{superatom} J. Zeiher, P. Schau\ss, S. Hild, T. Macr\`i, I. Bloch, and C. Gross, ``Microscopic Characterization of Scalable Coherent Rydberg Superatoms," Phys. Rev. X {\bf5}, 031015 (2015).






\bibitem{Lindblad} G. Lindblad, ``On the generators of quantum dynamical semigroups," Commun. Math. Phys. {\bf48}, 119 (1976).

\bibitem{noise} S. de L\'es\'eleuc, D. Barredo, V. Lienhard, A. Browaeys, and T. Lahaye, ``Analysis of imperfections in the coherent optical excitation of single atoms to Rydberg states," Phys. Rev. A. {\bf97}, 053803 (2018).

\bibitem{noise few atom} W. Lee, M. Kim, H. Jo, Y. Song, and J. Ahn, ``Coherent and dissipative dynamics of entangled few-body systems of Rydberg atoms,"  Phys. Rev. A. {\bf99} 043404 (2019)

\bibitem{Endres2020} I. S. Madjarov, J. P. Covey, A. L. Shaw,  J. Choi, A. Kale, A. Cooper, H. Pichler, V. Schkolnik, J. R. Williams, and M. Endres, ``High-fidelity entanglement and detection of alkaline-earth Rydberg atoms,"  Nat. Phys. {\bf16}, 857–861 (2020).




\bibitem{Schymik2021} K.-N. Schymik, S. Pancaldi, F. Nogrette, D. Barredo, J. Paris, A. Browaeys, and T. Lahaye, ``Single Atoms with 6000-Second Trapping Lifetimes in Optical-Tweezer Arrays at Cryogenic Temperatures," Phys. Rev. Applied {\bf 16}, 034013 (2021).



\end{thebibliography}
\end{document}